\documentclass{taira}
\usepackage{tabularx}
\usepackage{graphicx}
\usepackage[square,sort,comma,numbers]{natbib}
\usepackage{pifont}
\usepackage{soul}
\usepackage{amsmath}
\usepackage{amssymb}
\usepackage{amsfonts}
\usepackage{color}
\usepackage{mathtools}
\usepackage{relsize}
\usepackage{wasysym}

\newcommand{\vol}{\mathop{\ooalign{\hfil$V$\hfil\cr\kern0.08em--\hfil\cr}}\nolimits}

\newcommand{\bs}[1]{\boldsymbol{#1}}

\newcommand{\abs}[1]{{\left\lvert#1\right\rvert}}
\newcommand{\norm}[1]{{\left\lVert#1\right\rVert}_2}

\newcommand{\mrm}[1]{\mathrm{#1}}
\title[Phase-locking of laminar wake to periodic cylinder vibrations]{Phase-locking of laminar wake to periodic vibrations of a circular cylinder}

\author[M. A. Khodkar, J. T. Klamo and K. Taira]{M. A. Khodkar$^1$, Joseph T. Klamo$^2$ and Kunihiko Taira$^{1}$\thanks{Email addresses for correspondence: mkhodkar@ucla.edu}}

\affiliation{$^1$Department of Mechanical and Aerospace Engineering, University of California, Los Angeles, CA 90095, USA\\
$^2$Department of Systems Engineering, Naval Postgraduate School, Monterey, CA 93943, USA} 

\date{\today}

\pubyear{2020}
\begin{document}

\maketitle

\begin{abstract}

Phase synchronization between the vortex shedding behind a two-dimensional circular cylinder and its vibrations is investigated using the phase-reduction analysis. Leveraging this approach enables the development of a one-dimensional, linear model with respect to the limit-cycle attractor of the laminar wake, which accurately describes the phase dynamics of the high-dimensional, nonlinear fluid flow and its response to rotational, transverse and longitudinal vibrations of the cylinder. This phase-based model is derived by assessing the phase-response and sensitivity of the wake dynamics to impulse perturbations of the cylinder, which can be performed in simulations and experiments. The resulting model in turn yields the theoretical conditions required for phase-locking between the cylinder vibrations and the wake. We furthermore show that this synchronization mechanism can be employed to stabilize the wake and subsequently reduce drag. We also uncover the circumstances under which the concurrent occurrence of different vibrational motions can be used to promote or impede synchronization. These findings provide valuable insights for the study of vortex-induced body oscillations, the enhancement of aerodynamic performance of flyers, or the mitigation of structural vibrations by synchronizing or desynchronizing the oscillatory motions of body to the periodic wake.  

\end{abstract}

\begin{keywords}
phase reduction, synchronization, vortex-induced vibration (VIV).
\end{keywords}

%%%%%%%%%%%%%%%%%%%%%%%%%%%%%%%%%%%%% Introduction %%%%%%%%%%%%%%%%%%%%%%%%%%%%%%%%%%%%%
\section{Introduction \label{Sec:Intro}}

Understanding the dynamics of the wake behind a bluff body immersed in a moving fluid is critical for predicting the fluctuating pressure field and the resulting forces on the body.  By manipulating the wake directly behind an object, one can decrease drag on a body, increase lift on a foil, or improve energy harvesting of the system. Moreover, the characteristics of the downstream wake can influence mixing performance and increase the reaction rate of combustion processes.

The oscillating wake behind a cylinder in a moving fluid was first studied by Strouhal in 1878 ~\citep{Strouhal1878}. However, it was not until the late 1960s that researchers investigated the effects of an actively moving cylinder on the synchronization of shed vortices in the wake.  These studies pulsed a piezoelectric material positioned between two hemispheres \citep{Wehrmann1965}, forced a circular cylinder to move transverse to the flow \citep{Koopmann1967}, and allowed a circular cylinder to freely vibrate normal to the flow \citep{Ferguson1965}. Each of these different excitation methods caused the vortex shedding frequency to synchronize with the frequency of the cylinder motion when the frequency of that motion was close to the natural shedding frequency of the flow. These investigations confirmed that the controlled excitation of a cylinder at frequencies close to the vortex shedding frequency for a stationary cylinder resulted in important changes to the wake. An enormous amount of research has since been carried out exploring various aspects of forced and free vibrations of cylinders in free stream. These results are described in the numerous review articles \citep{Sarpkaya1979, Griffin1982, Bearman1984, Parkinson1989, Rockwell1994, Sarpkaya2004, Williamson2004} on forced and free vibration of cylinders. An important finding from these investigations was the synchronization of the wake shedding frequency to the oscillating frequency of the cylinder. The term ``lock-in'' was coined to describe such behavior.

Subsequent research has enhanced our understanding of synchronization and led to important insights. A flow visualization study \citep{Williamson1988} identified the various wake structures behind an oscillating cylinder and the approximate location of the synchronization boundaries for Re = 1,000. Perhaps the most important finding of their work was the observation that the wake can change significantly at frequencies not necessarily ``close'' to the natural shedding frequency.  This study revealed the existence of a much larger synchronization region than initially understood.  Moreover, the study reported the observation of two different wake structures within the synchronization region.  In an ensuing study, a dense parametric sweep of the amplitude and frequency space of a forced oscillating cylinder was completed to precisely identify the synchronization boundaries and two wake structures at $\mrm{Re}=4,000$ and $12,000$ \citep{Morse2009}.

Despite 50 years of research involving forced and free cylinder vibrations, there has been no technique to predict the boundaries of the synchronization region in the frequency-amplitude plane. Instead, we must laboriously find it experimentally or numerically through an extensive parametric study for a particular Reynolds number and cylinder shape of interest. 
To address this shortcoming, we put forth a phase-based analysis capable of identifying the synchronization region.  This approach only requires a limited set of experimental or numerical results to identify the phase sensitivity to various perturbations. The approach was previously used on the phase response to external excitation of low-Reynolds-number laminar wakes \citep{Taira2018, Khodkar2020}. In our study, we apply the phase-reduction analysis to flows in which the body is forced to oscillate in a rotary, crossflow (transverse) or streamwise (inline) fashion. We develop a linear, one-dimensional model with respect to the limit cycle (periodic flow) that accurately characterizes the conditions necessary for synchronization between the wake shedding frequency and the periodic cylinder vibrations. This model can further be utilized to examine how the concurrent presence of different vibrational modes, for example simultaneous rotary and crossflow vibrations, might allow for widening or narrowing the synchronization region.

Identification and prediction of the synchronization region is the critical aspect of vortex-induced vibration (VIV). This is because the free vibration of a structure specifically occur when the wake is inside the synchronization region. The importance of the vortex-induced vibration of structures for a range of engineering problems has been identified in the past few decades. The practical applications of VIV can be commonly seen in offshore platforms, heat exchangers, transmission lines and marine cables \citep{King1974, Rockwell1994, Williamson2004, Sarpkaya2004}.

The organization of the paper is as follows. The phase-based model is developed in Sec.~\ref{Sec:Theory}. Section \ref{Sec:DNS} begins with a short description of the numerical solver used for conducting the direct numerical simulations (DNS) of the flow.  We further present the phase properties of the flow for various types of cylinder oscillation, obtained by introducing impulsive perturbations of the cylinder at several times over a period in the DNS code. Section \ref{Sec:Results1} renders the model predictions for synchronization regions, and shows their accuracy by comparing them to the present DNS results as well as earlier experimental and numerical data. It also discusses the effect of phase-locking between the vortex shedding and cylinder oscillation on the wake structure and, consequently, lift and drag. Section \ref{Sec:Results2} demonstrates how the synchronization region of the periodic flow can be altered by superimposing different vibrational motions. Section \ref{Sec:Conclusion} concludes the paper with the summary of principal findings and their practical implications.

%%%%%%%%%%%%%%%%%%%%%%%%%%%%%%%%%%%%% Theory %%%%%%%%%%%%%%%%%%%%%%%%%%%%%%%%%%%%%
\section{Phase reduction of periodic flows \label{Sec:Theory}}

Let us consider the dynamics of an incompressible flow described by
\begin{equation}
\dot{\bs{q}} = \bs{N}(\bs{q}) \, , \label{Eqn:Dyn1}
\end{equation}
where $\bs{q}$ is the state vector of the flow. In this paper, we focus on the dynamics with respect to a stable limit cycle $\bs{q_0}$, for which the flow is periodic, satisfying $\bs{q_0}(\bs{x}, t+T) = \bs{q_0}(\bs{x}, t)$. Here, $T$ denotes the periodicity, which is related to the natural frequency $\omega_n$ by $\omega_n = 2\pi/T$. Due to the periodic nature of this base flow, its phase dynamics can be represented by  
\begin{equation}
\dot{\theta} = \omega_n \, , \quad \theta \in [-\pi, \pi] \, , \label{Eqn:Phase1}
\end{equation}
where $\theta$ is the phase of the limit-cycle oscillation. When the state does not reside on the limit-cycle attractor, but instead lies in its basin, the phase dynamics can be characterized by the phase function $\mathit{\Theta}(\bs{q})$ such that $\theta = \mathit{\Theta}(\bs{q})$. The use of $\mathit{\Theta}(\bs{q})$ in Eq.~(\ref{Eqn:Phase1}) yields  
\begin{equation}  
\dot{\theta} = \mathit{\dot{\Theta}}(\bs{q}) = \bnabla_{\bs{q}} \mathit{\Theta}(\bs{q}) \cdot \dot{\bs{q}} = \bnabla_{\bs{q}} \mathit{\Theta}(\bs{q}) \cdot \bs{N}(\bs{q}) = \omega_n \, . \label{Eqn:Phase2}
\end{equation}
Here, the phase function $\mathit{\Theta}(\bs{q})$ extends the definition of phase in the vicinity of the limit cycle.

Now, consider the introduction of a weak perturbation to the fluid system of Eq.~(\ref{Eqn:Dyn1})
\begin{equation}
\dot{\bs{q}} = \bs{N}(\bs{q})  + \epsilon \bs{f}(t) \, , \label{Eqn:Dyn2}
\end{equation}
with $\epsilon \ll 1$ and $\norm{\bs{f}} = 1$. This equation can be rewritten in terms of the phase function to obtain 
\begin{equation}  
\dot{\theta} = \mathit{\dot{\Theta}}(\bs{q}) = \bnabla_{\bs{q}} \mathit{\Theta}(\bs{q}) \cdot \dot{\bs{q}} = \bnabla_{\bs{q}} \mathit{\Theta}(\bs{q}) \cdot [\bs{N}(\bs{q}) + \epsilon \bs{f}(t) ] \, , \label{Eqn:Phase_forced}
\end{equation}
which along with Eq.~(\ref{Eqn:Phase2}) gives
\begin{equation}  
\dot{\theta} = \omega_n + \epsilon \bnabla_{\bs{q}} \mathit{\Theta}(\bs{q})\big|_{\bs{q} = \bs{q_0}}\cdot\bs{f}(t) \, , \label{Eqn:LRF}
\end{equation}
where we have neglected the higher-order terms and derived a linear model with respect to the limit cycle $\bs{q_0}(t)$. This equation describes the phase dynamics of the nonlinear and high-dimensional flow only via a scalar phase variable. In the rest of the paper, $\bnabla_{\bs{q}} \mathit{\Theta}(\bs{q}) |_{\bs{q} = \bs{q_0}}$ is denoted by $\bs{Z}(\theta)$ and is called the phase-sensitivity function.

The focus of the present work is to develop a general framework which can be implemented both numerically and experimentally. Therefore, adjoint-based approaches requiring the explicit knowledge of governing equations \citep{Ermentrout2010, Nakao2016} are not pursued, and instead, a direct method, which evaluates $\bs{Z}$ by applying weak impulse perturbations at different times (phases) over a period, is employed. These impulses are in the form $I \delta(t - t_0) \hat{\bs{e}}_{\bs{j}}$, where $I$, $\delta(t-t_0)$ and $\hat{\bs{e}}_{\bs{j}}$ respectively indicate the perturbation amplitude, a Dirac delta function centered at $t_0$ and the unit vector in the direction of the impulse. The delta function $\delta(t-t_0)$ is modeled as the following narrow Gaussian function in the numerical solver 
\begin{equation}  
\delta(t-t_0) = \frac{1}{\sqrt{2\pi} \sigma} \exp\bigg[ - \frac{1}{2}\Big(\frac{t - t_0}{\sigma}\Big)^2 \bigg] \, , \label{Eqn:Dirac_delta}
\end{equation}
where $\sigma = 10 \Delta t$, with $\Delta t$ being the DNS time step (Sec.~\ref{Sec:DNS}). 

After sufficiently long time, the transient effects of adding impulse perturbation to the periodic flow vanish, and the state of the flow relaxes to its limit cycle. However, the impulse creates an asymptotic phase shift, referred to as the phase-response function $g(\theta; I\hat{\bs{e}}_{\bs{j}})$, which can be measured for any given phase $\theta$ at which the flow is initially perturbed. The phase-sensitivity function $\bs{Z}$ is then calculated as       
\begin{equation}  
Z_j(\theta) = \lim_{I \rightarrow 0} \frac{g(\theta; I\hat{\bs{e}}_{\bs{j}})}{I} \approx \frac{g(\theta; I\hat{\bs{e}}_{\bs{j}})}{I} \, , \label{eqn:PSF}
\end{equation}     
assuming that $I \ll 1$. Repeating the same procedure for different values of $\theta$ over the entire period provides the $2\pi$-periodic phase-sensitivity function \citep{Nakao2016, Taira2018}. 

Synchronization or phase-locking between the periodic flow and any harmonic oscillators (e.g., external actuations and forced or flow-induced body vibrations) can be achieved if and only if the temporal change $\dot{\phi}$ in the phase difference between the flow and the forced oscillator ($\phi = \theta(t) - \omega_f t$) approaches zero, while the oscillator's frequency $\omega_f$ is not very different from $\omega_n$ or its harmonics. Substituting $\dot{\theta}$ from Eq.~(\ref{Eqn:LRF}) enables expressing this synchronization condition in the following analytical form
\begin{equation}  
\dot{\phi} = \epsilon[\Delta + \bs{Z}(\phi + \omega_f t) \cdot \bs{f}(t)] \rightarrow 0 \, , \label{Eqn:Relative_phase1}
\end{equation}
where $\Delta = (\omega_n - \omega_f)/\epsilon = \mathcal{O}(1)$. 

Since $\dot{\phi} \ll \omega_f$, the right-hand side of equation (\ref{Eqn:Relative_phase1}) can be approximated by its mean over one period \citep{Kuramoto1984, Ermentrout2010}, leading to   
\begin{equation}  
\dot{\phi} = \epsilon[\Delta + \mathit{\Gamma}(\phi)]  \, , \label{Eqn:Relative_phase2}
\end{equation}
where the $2\pi$-periodic phase-coupling function $\mathit{\Gamma}(\phi)$ is formulated as 
\begin{equation}  
\mathit{\Gamma}_m(\phi) = \frac{\omega_f}{2\pi} \int_{-\pi/\omega_f}^{\pi/\omega_f} \bs{Z} (\phi + \omega_f \tau/m ) \cdot \bs{f}(\tau) \mrm{d}\tau = \frac{1}{2\pi} \int_{-\pi}^{\pi} \bs{Z}(\phi + \varphi/m) \cdot \bs{f}(\varphi) \mrm{d}\varphi \, , \label{Eqn:PCF}
\end{equation}
with $m \in \mathbb{N}$ indicating the number of harmonic at which the synchronization is investigated. For the $m{\mrm{th}}$ harmonic, synchronization to the frequency $m\omega_n$ can be studied. Stable phase-locking between the oscillator and the periodic flow requires
\begin{equation}  
\epsilon \mathit{\Gamma}_{\mrm{min}} < \omega_f/m - \omega_n < \epsilon \mathit{\Gamma}_{\mrm{max}} \, , \label{Eqn:Sync_condition}
\end{equation}
where the minimum and maximum values of $\mathit{\Gamma}$ are denoted by $\mathit{\Gamma}_{\mrm{max}}$ and $\mathit{\Gamma}_{\mrm{min}}$, respectively. Note that, in this study, phase-locking at higher harmonics ($m > 1$) is always subharmonic, as will be discussed in Sec.~\ref{Sec:DNS_Results}. In a subharmonic synchronization, the periodc flow is synchronized to $\omega_f/m$ instead of $\omega_f$. Equation~(\ref{Eqn:Sync_condition}) suggests that the region of synchronization in the frequency-amplitude plane, also known as the Arnold tongue \citep{Arnold1997}, grows with $\mathit{\Gamma}_{\mrm{max}} - \mathit{\Gamma}_{\mrm{min}}$. The synchronizability parameter $S \equiv \mathit{\Gamma}_{\mrm{max}} - \mathit{\Gamma}_{\mrm{min}}$ is thus defined as a measure to quantify how easily phase-locking can occur.

%%%%%%%%%%%%%%%%%%%%%%%%%%%%%%%%%%%%% DNS and Results %%%%%%%%%%%%%%%%%%%%%%%%%%%%%%%%%%%%%
\section{Phase synchronization of wake flow to cylinder oscillations \label{Sec:DNS_Results}}

We explore the influence of fundamental vibrational motions of the cylinder on the synchronization properties of the wake flow. This section commences with a brief description of the DNS solver used for simulating the 2D incompressible flow over the oscillating cylinder. It further presents the phase-sensitivity and phase-coupling functions obtained using perturbed DNS simulations. In the latter part of the section, the theoretical and numerical Arnold tongues are compared, and the possibility of stabilizing the wake via cylinder vibrations is investigated. The section is concluded by discussing the impact of combining different types of oscillation on the synchronizability of the periodic wake.   

\subsection{Numerical solver \label{Sec:DNS}}

\begin{figure}
 \centerline{\includegraphics[width=1.\textwidth]{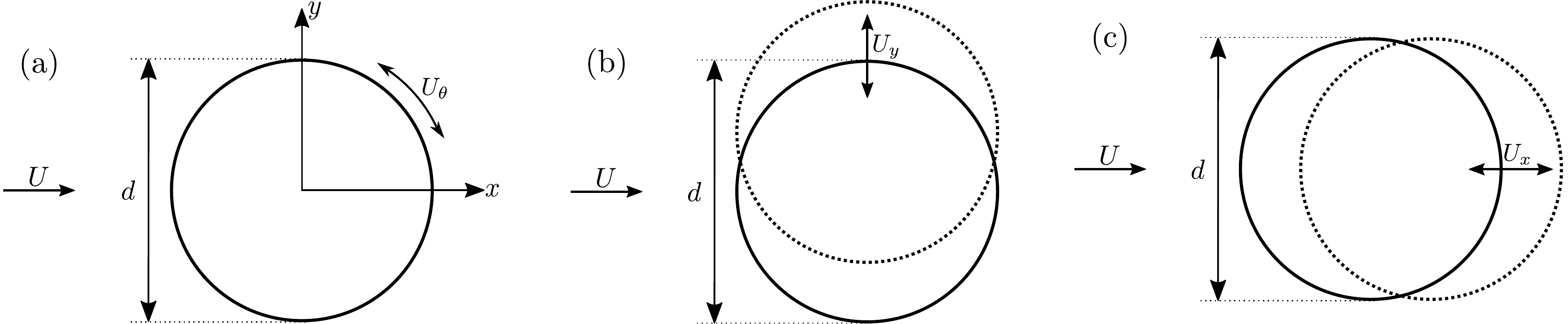}}
\caption{Schematic of a uniform flow past a 2D circular cylinder, when the cylinder oscillates in a (a) rotary, (b) streamwise or (c) crossflow fashion, with the respective tangential, streamwise or crossflow velocities $U_{\theta}$, $U_x$ or $U_y$. All these velocities are given by $U_f \sin(\omega_f t)$, with $U_f$ and $\omega_f$ indicating the velocity oscillation amplitude and frequency, respectively. %(d) Lift coefficient $C_L$ as a function of phase value $\theta$.
}
\label{Fig:Setup}
\end{figure}

The dimensionless governing equations of a 2D incompressible flow are 
\begin{eqnarray}
\frac{\partial \bs{u}}{\partial t} + \bs{u} \cdot \bnabla{\bs{u}} = - \bnabla p + \frac{1}{\mrm{Re}} \nabla^2 \bs{u}   \,  \quad \textrm{and} \quad \bnabla \cdot \bs{u} = 0 \, .
\label{Eqn:DNS}
\end{eqnarray}
Here, $\bs{u}$ and $p$ represent the nondimensional velocity and pressure fields, respectively, while the cylinder diameter $d$ and the freestream velocity $U$ are selected as the characteristic length and velocity (Fig.~\ref{Fig:Setup}). The Reynolds number is defined as $\mrm{Re} \equiv Ud/\nu$, where $\nu$ is the kinematic viscosity. In all considered cases, we set $\mrm{Re} = 100$. The lift coefficient $C_L$ and its time derivative $\dot{C}_L$ are used to determine the phase with $\theta = \tan^{-1}(C_L/\dot{C}_L)$.

%*******************************************************************
% Figure: Configuration of the problem
%*******************************************************************

%*******************************************************************
% End Figure
%******************************************************************* 

The immersed boundary projection method \cite{Taira2007} is utilized to conduct the DNS of the flow, with the multi-domain technique \cite{Colonius2008} employed for far-field boundary conditions. The center of the cylinder is located at the origin, and the computational domain is stretched over $(x/d, y/d) \in [-39.5, 40.5] \times [-40, 40]$. The horizontal and vertical spacings of the finest domain, which is the closest to the cylinder, are uniform with $\Delta x/d = \Delta y/d = 0.025$. The time step $\Delta t = 0.005$ satisfies $U \Delta t/ \Delta x < 0.5$. The natural shedding frequency $\omega_n$, the aerodynamic forces on the cylinder, and the velocity and vorticity fields provided by the unperturbed computational setup of this section agree closely with those of previous works \citep{Munday2013, Taira2018}.

%************************************************************************
% Figure: Phase properties of all oscillation cases 
%************************************************************************
\begin{figure}
  \centerline{\includegraphics[width=1.\textwidth]{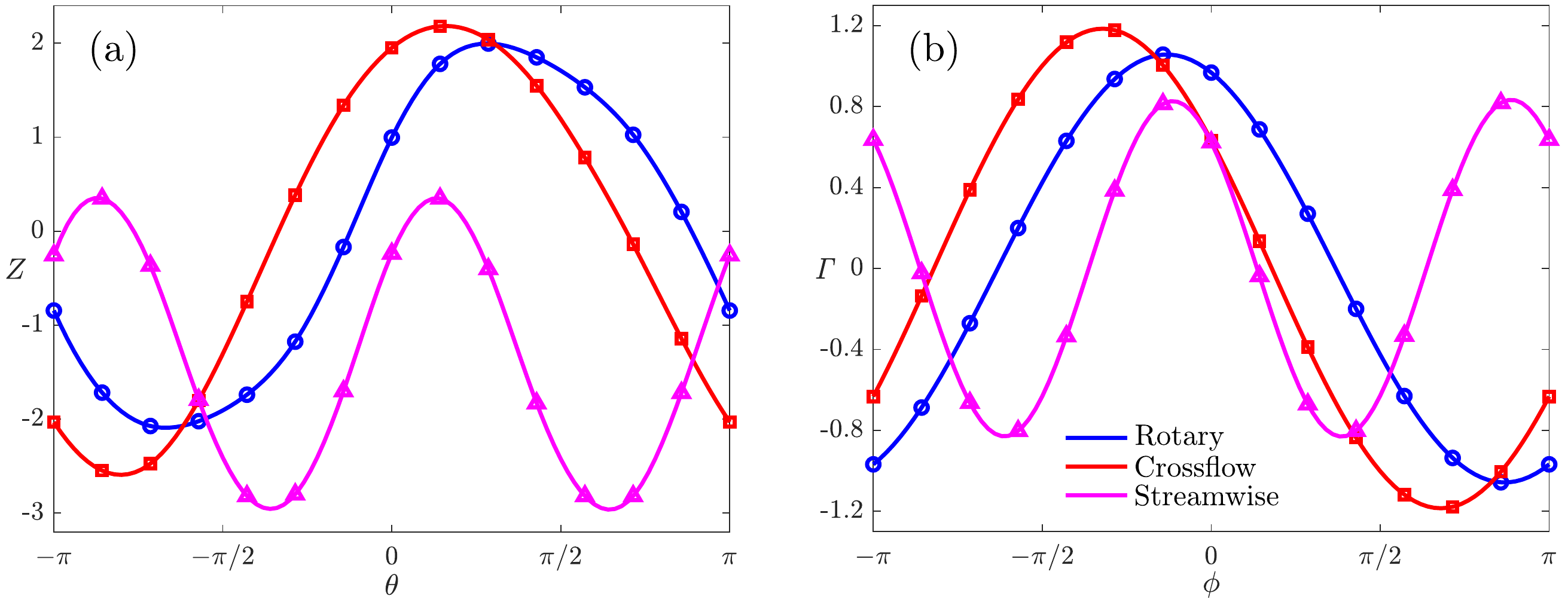}}
  \caption{(a) Phase-sensitivity $\mathit{Z}$, and (b) phase-coupling functions $\mathit{\Gamma}$ for different types of cylinder oscillation. Note that presented $\mathit{\Gamma}$ for the rotary and crossflow oscillations are the first-harmonic phase-coupling functions, while that of streamwise oscillation is calculated at the second harmonic.
}
\label{Fig:Z_Gamma}
\end{figure}
%************************************************************************
% End Figure
%************************************************************************ 

To determine the phase-sensitivity function of rotary oscillation, the cylinder is impulsively moved with the tangential velocity $U_{\theta} = U_f \delta(t-t_0)$ (Fig.~\ref{Fig:Setup}(a)) at different phases over a period. For the crossflow and streamwise oscillations, body oscillations are incorporated in the reference frame of the body by subtracting the cylinder speed $U_x = U_f \delta(t-t_0)$ or $U_y = U_f \delta(t-t_0)$ from the freestream velocity (Figs.~\ref{Fig:Setup}(b) and (c)). These impulses are introduced to the governing equations as accelerations to the reference frame of the body, and are numerically enforced by modifying the boundary conditions. The phase-sensitivity function associated with each case is then calculated by determining the asymptotic phase shift in the time series of $C_L$, and is depicted in Fig.~\ref{Fig:Z_Gamma}(a). Lift and drag are evaluated in the reference frame moving with the cylinder. We also remark that the linearity assumption of Sec.~\ref{Sec:Theory} necessitates choosing $U_f \leq 0.05$ ($U_f \leq 0.025$) for rotary and crossflow oscillations (streamwise oscillation). 

Once the phase-sensitivity function $Z$ is found, the corresponding phase-coupling function $\mathit{\Gamma}$ is obtained from Eq.~(\ref{Eqn:PCF}), and is displayed in Fig.~\ref{Fig:Z_Gamma}(b). Although the phase-sensitivity functions of rotary and crossflow oscillations peak at different phase values, their maximum and minimum values are fairly close. In addition, both oscillations yield similar values for the synchronizability at the first harmonic ($S = 2.11$ and $2.37$, respectively for the rotary and crossflow oscillations) and very small values for the synchronizability at the second harmonic ($S = \mathcal{O}(0.01)$ for both), confirming the observations of former studies that rotary and transverse vibrations only exhibit first-harmonic synchronization \citep{Tokumaru1991, Placzek2009}. The $\pi$-periodicity of the phase-sensitivity function for the steamwise oscillation is also consistent with the findings of earlier investigations \citep{Tanida1973, Mdallal2007}, demonstrating that the wake flow synchronizes to the inline oscillation only subharmonically at the second harmonic, as this property results in $S \approx 0$ at the first harmonic, and much larger values for synchronizability ($S = 1.66$) at the second harmonic.

\subsection{Synchronization and force properties \label{Sec:Results1}}

%**************************************************************************************************
% Figure: Arnold tongues and time-averaged drag coefficients
%**************************************************************************************************
\begin{figure}
  \centerline{\includegraphics[width=1.\textwidth]{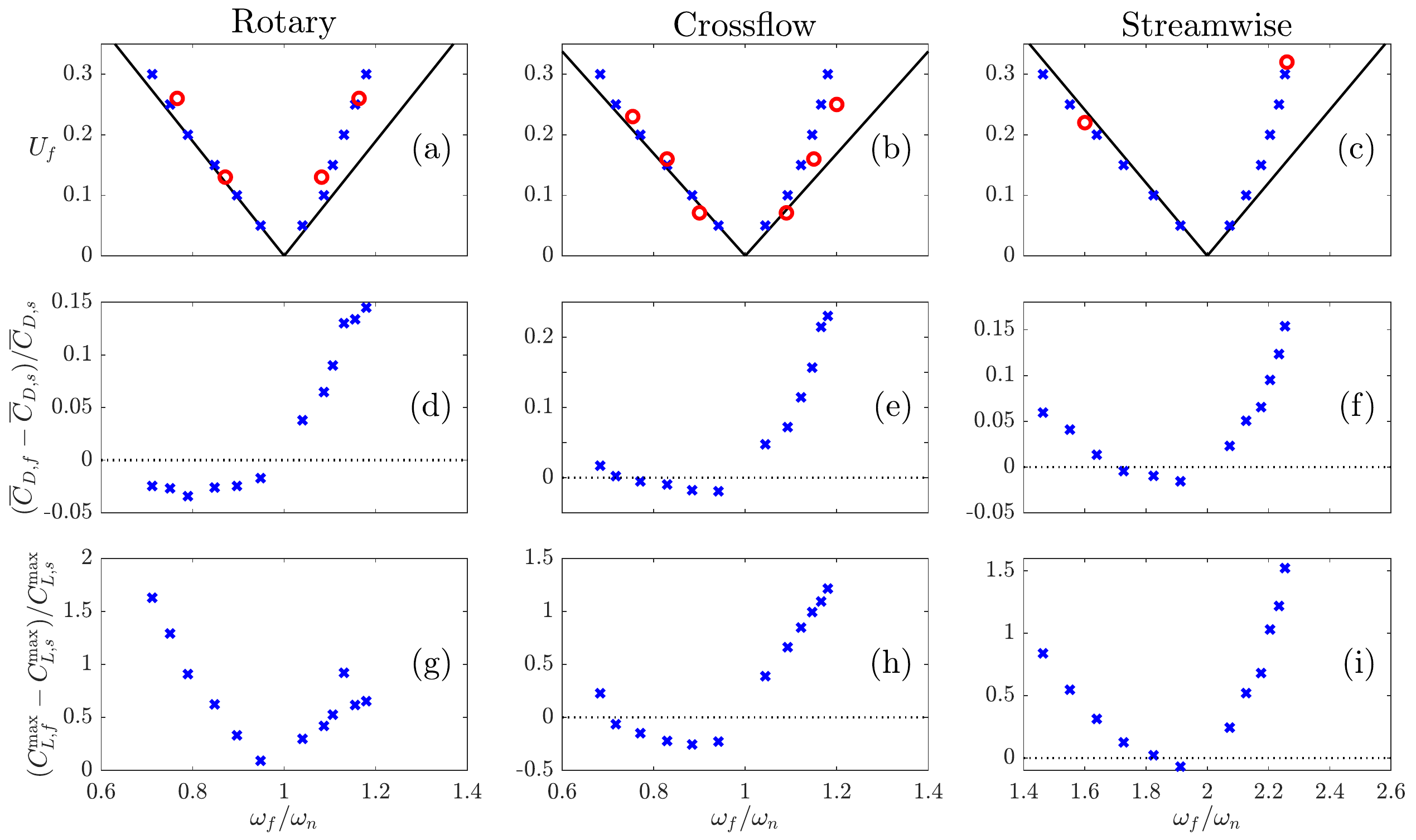}}
  \caption{Top: Theoretical and numerical boundaries of synchronization for (a) rotary, (b) crossflow and (c) streamwise oscillations in the frequency-amplitude plane, indicated by black solid lines and blue crosses, respectively. Middle: Time-averaged drag coefficients $\overline{C}_D$ for the synchronized cases of (d) rotary, (e) crossflow and (f) streamwise oscillations. Bottom: Same as middle but for maximum lifts $C_L^{\mrm{max}}$. Subscripts $f$ and $s$ refer to the oscillatory and stationary cylinder cases, respectively. Horizontal dotted lines represent zero values, or when the cylinder is at rest. Red circles of panels (a) to (c) show the numerical results of \cite{Baek1998} at $\mrm{Re} = 110$, the experimental observations of \cite{Koopmann1967} at $\mrm{Re} = 100$, and the findings of \cite{Tanida1973} at $\mrm{Re} = 80$, respectively.}
\label{Fig:Arnold_tongues}
\end{figure}
%**************************************************************************************************
% End Figure
%**************************************************************************************************   

%**************************************************************************************************
% Figure: Contours of vorticity field for different oscillatory cases 
%**************************************************************************************************
\begin{figure}
  \centerline{\includegraphics[width=1.0\textwidth]{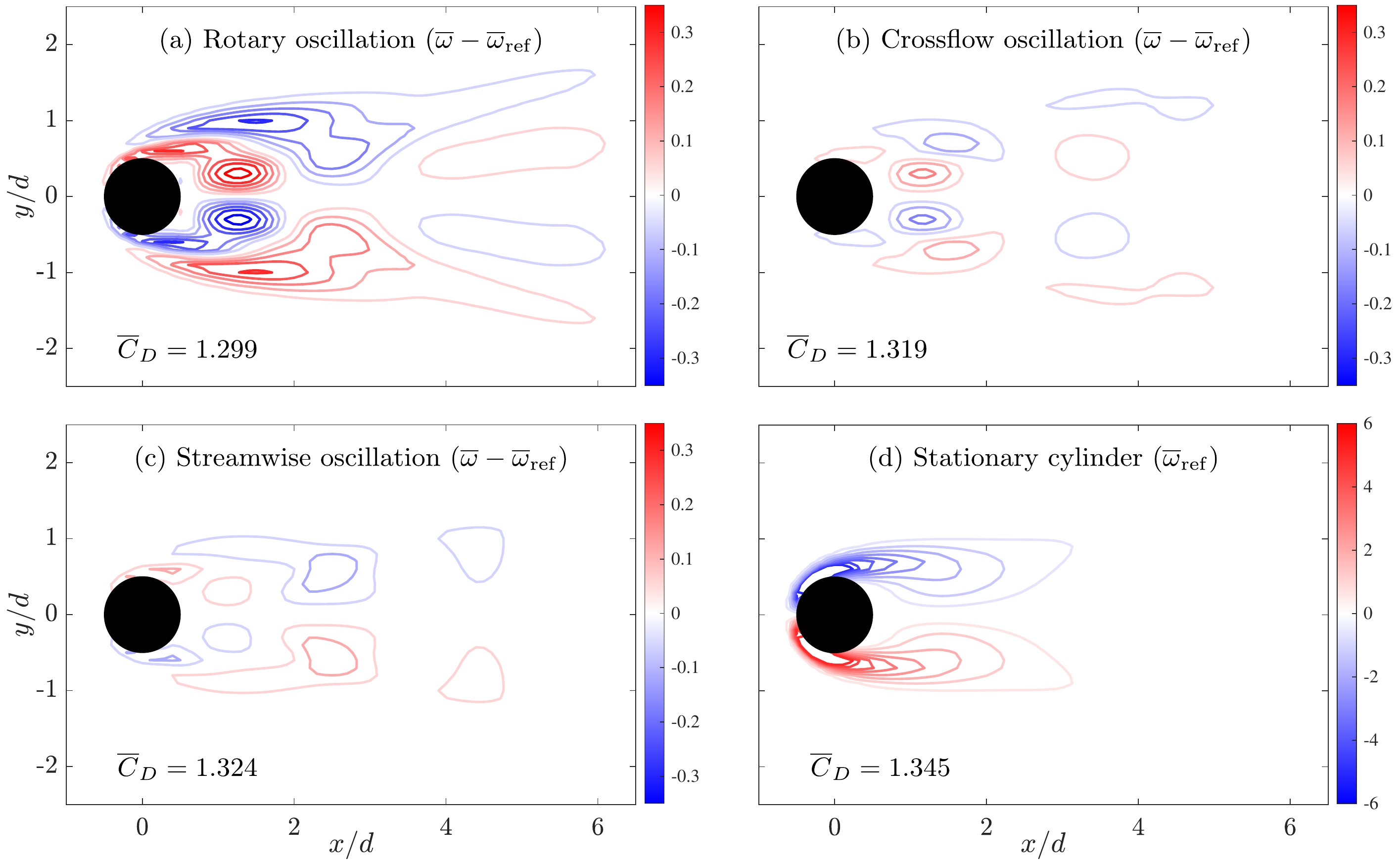}}
  \caption{(a-c) Difference between the time-averaged vorticity fields of the synchronized cases with the lowest mean drag coefficient $\overline{C}_D$ ($\overline{\omega}$) and that of the stationary cylinder ($\overline{\omega}_\mrm{ref}$). (d) Time-averaged vorticity field of the base flow.}
\label{Fig:VF}
\end{figure}
%**************************************************************************************************
% End Figure
%**************************************************************************************************

The Arnold tongues (solid lines) revealed by the phase-reduction analysis for rotary, crossflow and streamwise oscillations are compared against the corresponding boundaries of synchronization obtained from DNS (blue crosses) in Fig.~\ref{Fig:Arnold_tongues}, for which $U_{\theta}$, $U_x$ or $U_y$ are given by $U_f \sin(\omega_f t)$. While synchronization to the natural shedding frequency is considered for the rotary and crossflow vibrations, the synchronization of the streamwise oscillation is studied at its second harmonic. The blue crosses correspond to the oscillation frequencies that maximize $\abs{\omega_f - m\omega_n}$ for a specified value of $U_f$ while maintaining the difference between the shedding frequency and $\omega_f$ below $1\%$. The red circles are the results extracted from previous studies \citep{Baek1998, Koopmann1967, Tanida1973}, with $\mrm{Re}$ varying from 80 to 110. The left branches of the present numerical and theoretical boundaries of synchronization are in good agreement up to large values of velocity oscillation amplitude ($U_f \lesssim 0.25$). In contrast, on the right branch of Arnold tongues, agreement is limited to moderate oscillation amplitudes ($U_f \lesssim 0.1$). As $U_f$ gradually increases, the state of the flow departs farther from its stable limit cycle and, as a result, nonlinear effects become progressively more manifest. It has also been noted by earlier investigations that delaying the vortex shedding process by phase-locking it to oscillation frequencies below $\omega_n$ uses less actuation energy compared to when the shedding is advanced, assuming that the desired frequency shift is identical for both, as the latter requires larger oscillation amplitudes \citep{Koopmann1967, Barbi1986, Baek1998}. This is of aerodynamic importance, since delaying vortex shedding is the primary mechanism for reducing the wake unsteadiness and, subsequently, lowering drag at the Reynolds number and frequencies under consideration, whereas accelerating the shedding process affects the aerodynamic performance by exciting the shear-layer instability only at higher Reynolds numbers ($\mrm{Re} \gtrsim 500$) and frequencies ($\omega_f \gtrsim 4\omega_n$) \citep{Filler1991}. It is thus encouraging to find that the aerodynamically advantageous strategy needs smaller amplitude for control inputs, and can be accurately predicted using the present phase-based model. Among all three oscillatory modes studied in the current work, the rotary oscillation leads to the largest drag reduction, by around $4\%$ decrease with respect to the stationary cylinder case (Figs.~\ref{Fig:Arnold_tongues}(d-f)). The difference between the time-averaged vorticity fields of the synchronized wake flows of Fig.~\ref{Fig:Arnold_tongues} minimizing $\overline{C}_D$ and that of the base flow (stationary cylinder case) are portrayed in Fig.~\ref{Fig:VF} for all three oscillatory motions. As shown in this figure, a lower value for $\overline{C}_D$ corresponds to a more elongated and streamlined wake caused by delayed vortex shedding. Maximum lift coefficients $C_L^{\mrm{max}}$ can also be altered by cylinder vibration, as shown in Figs.~\ref{Fig:Arnold_tongues}(g-i). Crossflow oscillation noticeably reduces the amplitudes of lift when the vortex shedding is phase-locked to $0.75 \lesssim \omega_f/\omega_n \lesssim 1$. The synchronized cases on the right branch of Arnold tongues augment $C_L^{\mrm{max}}$, with crossflow and streamwise oscillations being more effective.

\subsection{Concurrent vibrations \label{Sec:Results2}}

In real-world examples, structural vibrations or oscillatory motions of flyers are typically a combination of the aforementioned vibrational modes \citep{Rockwell1994, Williamson2004}. It is thus practically more important to analyze scenarios in which all these motions occur simultaneously. Combining the streamwise oscillation with any of the other motions breaks the $\pi$-periodicity of the phase-sensitivity function and the symmetry of the inline oscillation with respect to the centerline of the wake flow, leading to no lock-on to the second harmonic. Furthermore, since the phase-coupling function of the sreamwise oscillation is nearly vanishing at the first harmonic, the synchronization properties of the flow at the this harmonic are negligibly affected by the presence of this oscillation. Hence, we confine our attention to the superposition of rotary and crossflow oscillations in what follows.

%**************************************************************************************************
% Figure: Contour of synchronizability of combined oscillations 
%**************************************************************************************************
\begin{figure}
  \centerline{\includegraphics[width=1.0\textwidth]{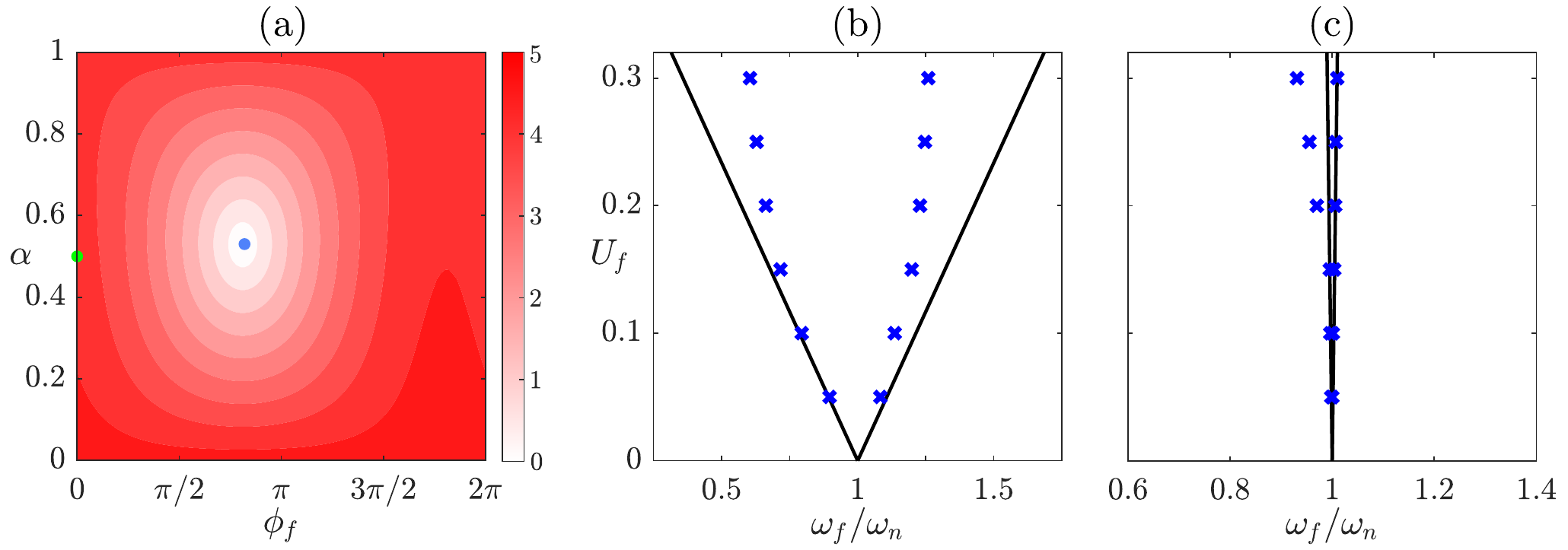}}
  \caption{(a) Synchronizability as functions of the weight $\alpha$ and phase difference $\phi_f$ of concurrent rotary and crossflow oscillations. (b) Theoretical (solid lines) and numerical (blue crosses) Arnold tongues corresponding to the green point of panel (a) for which $\alpha = 0.5$ and $\phi_f = 0$. (c) Same as panel (b) but for the blue point of panel (a) at which synchronizability is minimized.}
\label{Fig:Combined}
\end{figure}
%**************************************************************************************************
% End Figure
%**************************************************************************************************

Suppose that the cylinder oscillates rotationally and transversely at the same frequency, but with different amplitudes and phases. The combined motion can be mathematically expressed as       
\begin{equation}  
	\bs{f}(t) = 2U_f \big[ \alpha \sin(\omega_f t + \phi_f) \hat{\bs{e}}_{\bs{\theta}} +  (1 - \alpha) \sin(\omega_f t) \hat{\bs{e}}_{\bs{y}} \big] \, , \label{Eqn:Combined}
\end{equation} 
where $\phi_f$ is the phase difference between the rotary and crossflow vibrations, and the weight $\alpha$ varies from 0 to 1. The unit vectors $\hat{\bs{e}}_{\bs{\theta}}$ and $\hat{\bs{e}}_{\bs{y}}$ are oriented in the tangential and vertical directions, respectively. The synchronizability of the motion described by Eq.~(\ref{Eqn:Combined}) is demonstrated as functions of $\alpha$ and $\phi_f$ in Fig.~\ref{Fig:Combined}. The synchronizability is maximized when the cylinder motion is a purely crossflow vibration ($\alpha = 0$). The results of this case have been studied in detail in Sec.~\ref{Sec:Results1}. The blue circle of Fig.~\ref{Fig:Combined}(a) with $\alpha = 0.53$ and $\phi_f = 0.82\pi$ corresponds to the point at which synchronizability is the lowest ($S = 0.064$). The green circle represents the case at which the rotary and crossflow oscillations are added with the same amplitude and no phase difference ($\alpha = 0.5$ and $\phi_f = 0$). As expected, rotary and crossflow oscillations occurring simultaneously with the same amplitude and phase, amplify the synchronization properties of each other, as the phase-sensitivity functions of the two oscillations are nearly in phase (Fig.~\ref{Fig:Z_Gamma}(a)). Consequently, the resulting Arnold tongue (Fig.~\ref{Fig:Combined}(b)) is almost twice as wide wider compared to flows in which only one of the oscillations is present (Figs.~\ref{Fig:Arnold_tongues}(a) and (b)). This enables synchronization with frequencies much farther from $\omega_n$ for a fixed value of $U_f$. On the other hand, the proper choice for the amplitude and phase difference of the two oscillations can result in $S \approx 0$, as can be seen in Fig.~\ref{Fig:Combined}(c). However the numerical synchronization region is slightly wider than that predicted by the phase-based model (especially at large amplitudes), overall, the present model proves successful in proposing strategies for combining various vibrational motions that lead to the amplification or attenuation of the synchronization properties of the wake flow.

In a more comprehensive case, the two vibrations might not have the same frequency, which can give rise to a quasiperiodic flow with two dominant frequencies, if the cylinder wake synchronizes to both oscillations. Such case is beyond the scope of the present work, as it cannot be characterized by a single limit-cycle attractor, and therefore is not studied here.

In addition to enabling the analysis of realistic flows that incorporate complex structural vibrations or body oscillations, the present framework shows promise for designing optimal control strategies for achieving a desired aerodynamic target, e.g., an increased lift or a reduced drag, with the methodical superposition of different vibrational modes, or the introduction of external oscillators such as periodic actuations to the wake flow of a vibrating body. Furthermore, the resonance between the periodic flow and vortex-induced body vibrations can be prevented by substantially shrinking the synchronization region of the flow, as performed for the example studied in Sec.~\ref{Sec:Results2}, averting structural damages. As noted by \citet{Bearman1984}, flows past a forcibly oscillating body and a freely vibrating one are assumed to produce nearly identical patterns, so long as the amplitude and frequency of oscillations as well as the Reynolds number of the flows are the same. We thus reasonably expect that the general conclusions of this study hold for dynamics related to vortex-induced vibrations.

%%%%%%%%%%%%%%%%%%%%%%%%%%%%%%%%%%%%% Conclusions %%%%%%%%%%%%%%%%%%%%%%%%%%%%%%%%%%%%%
\section{Concluding Remarks \label{Sec:Conclusion}}

A detailed phase-based analysis of the response of a 2D periodic cylinder wake to excitations due to the cylinder vibrations was conducted. This analysis provided a theoretical approach for prediction of the required excitation amplitude and frequency to cause synchronization of the wake to the cylinder motion. The criteria for wake synchronization to rotary, crossflow and streamwise vibrations of the cylinder at $\mrm{Re} = 100$ were determined. A direct method was utilized which relies on impulsively perturbing the cylinder (or freestream velocity) at different times during the shedding cycle, to reveal the phase-response and sensitivity of the wake dynamics to the cylinder vibrations. This method avoids dealing with equations that capture the complex dynamics of the wake, and its numerical or experimental execution is straightforward and inexpensive. Our theoretical predictions for the synchronization region at $\mrm{Re} = 100$ for all three vibrational motions showed good agreement with a limited set of previous experimental data and a larger set of data generated by the present study. Finally, we demonstrated the possibility of widening or narrowing the synchronization region by combining the fundamental modes of vibration.

Although the authors anticipate that the general findings of the present work remain valid for practical applications of VIV that occur at higher Reynolds numbers, further work is needed for confirmation. We plan to examine the accuracy of the predicted synchronization boundaries using our phase-reduction analysis for each fundamental vibrational mode to experimentally determined ones at higher $\mrm{Re} = \mathcal{O}(10^3)$. This would confirm if the linear phase-based approach is sufficient to handle the complex and fully three-dimensional wakes emerging in actual VIV scenarios. If portions of the predicted synchronization boundaries at higher Reynolds numbers are not predicted accurately enough, such as the higher frequency side branch, then the inclusion of higher-order terms into the analysis is possible. These higher Reynolds numbers will also ensure that both the 2S and 2P wake structures, observed in previous studies \citep{Williamson1988, Carberry2001, Morse2009}, exist in different sections of the synchronization. This will allow us to demonstrate that, not only can we identify the synchronization region boundaries, but also the boundary between these two wake structures within the synchronization region can be detected using this phase-based methodology.

The phase-based model has remarkable potential once the accuracy of the approach at predicting the conditions leading to phase-locking between the vortex shedding and cylinder vibrations at higher Reynolds number can be established. A whole new direction of VIV research can be guided by our finding that the coupling of two different vibrational modes can either widen or narrow the synchronization region. A wider region allows for synchronization to be achieved using lower amplitudes which require less excitation. On the other hand, narrowing the region would provide another approach, other than using strakes, to suppress VIV. This can be attained through small, multiple-degree-of-freedom motions, to keep the wake unsynchronized and thereby substantially mitigating structural fatigue.

%%%%%%%%%%%%%%%%%%%%%%%%%%%%%%%%%%%%% Acknowledgements %%%%%%%%%%%%%%%%%%%%%%%%%%%%%%%%%%%%%
\section*{Acknowledgment \label{Sec:Acknowledgment}}

We gratefully acknowledge the support from the US Air Force Office of Scientific Research (Grant: FA9550-16-1-0650, Program Managers: Drs. Gregg Abate and Douglas Smith) and the Army Research Office (Grant: W911NF-19-1-0032, Program Manager: Dr. Matthew J. Munson).

% Create the reference section using BibTeX:

\bibliographystyle{taira}
\bibliography{Main}

\end{document}